\documentstyle[preprint,aps,epsfig]{revtex}
\begin{document}

\tighten
\draft
\preprint{
\vbox{
\hbox{August 1997}
\hbox{U.MD. PP\# 98-003}
\hbox{DOE/ER/40762-125}
\hbox{ADP-97-25/T260}
}}

\title{HERA Anomaly and Hard Charm in the Nucleon}
\author{W. Melnitchouk$^1$ and A. W. Thomas$^2$}
\address{$^1$	Department of Physics,
		University of Maryland,
		College Park, Maryland 20742, USA}
\address{$^2$	Department of Physics and Mathematical Physics,
		and Special Research Center for the
		Subatomic Structure of Matter,
		University of Adelaide,
		Adelaide 5005, Australia}

\maketitle

\begin{abstract}
We explore the possibility that the excess neutral and charged current
events seen by the H1 and ZEUS Collaborations at HERA at large $x$ and
$Q^2$ could arise from a hard charm component of the nucleon.
While the symmetric intrinsic charm hypothesis is unable to account
for the HERA anomaly, a non-symmetric charm distribution generated
non-perturbatively, for which $\overline c$ is much harder than $c$,
can produce significant enhancement of cross sections at the HERA
kinematics.
\end{abstract}
\pacs{PACS numbers: 13.60.Hb, 14.65.Dw, 12.38.Qk}

%%%%%%%%%%%%%%%%%%%%%%%%%%%%%%%%%%%%%%%%%%%%%%%%%%%%%%%%%%%%%%%%%%%%%%%%%%%%%%%%
The recent observation by the H1 \cite{H1} and ZEUS \cite{ZEUS} 
Collaborations at HERA of an excess of events at large $x$ and $Q^2$
in $e^+ p$ neutral current (NC) and charged current (CC) deep-inelastic 
scattering (DIS) has prompted numerous speculations about whether one
has seen evidence for physics beyond the standard model.
Most explanations have concentrated on leptoquarks or contact interactions
as the source for the anomalously large cross sections --- for a recent
overview of the theoretical discussions see Ref.\cite{REV}.
Using a simple, non-perturbative model for the intrinsic charm component
of the nucleon wave function, we find a hitherto unnoticed asymmetry,
with the $\overline c$ component significantly harder than $c$, which
has important consequences for the HERA NC and CC events.
Confirmation of such an asymmetry would challenge our understanding of
the non-perturbative structure of hadrons within QCD.

The ZEUS experiment collected only NC data, while H1 observed both NC
and CC DIS events.
The evidence for the excess was found in data with $Q^2 \agt 10 000$~GeV$^2$,
and with an invariant mass $M = \sqrt{x s}$ of the virtual photon---parton
system above $M \sim 150$~GeV, where $x$ is the Bjorken variable, and
$s \approx$ (300~GeV)$^2$ is the $e^+ p$ center of mass energy squared.
For $Q^2 > 35 000$~GeV$^2$ two events were observed in the ZEUS data
\cite{ZEUS} while $\sim 0.15$ were expected from the standard DIS model,
and for $x > 0.55$ and $y = M^2/Q^2 > 0.25$ four events were found with
$\sim 0.9$ anticipated.
For $Q^2 > 15 000$~GeV$^2$ the H1 data sample contained 12 NC (4 CC)
events, where $\sim 4.7$ (1.8) were expected.
On the face of it, this would amount to a probability of $\alt 1\%$ that the 
excess events arise from statistical fluctuations within the standard model.
On the other hand, it has been argued \cite{COMP} that the H1 and ZEUS
data sets are at present as incompatible with each other as each is with
the standard DIS model, so that more data analysis would be needed before
definitive conclusions could be reached.

Given the importance of the potential discovery of new physics, if the
excess persists in future HERA data it will be essential that one exhausts
all possibilities that the phenomenon can be accommodated within standard
model parameters.
Recently, Kuhlmann, Lai and Tung \cite{KLT} have suggested that an
increased valence $u$ quark distribution at large $x$ could enhance
the deep-inelastic cross sections appreciably.
The suggested modification to the valence $u$ distribution,
$u_V(x) \rightarrow u_V(x) + \delta u(x)$, where
\begin{eqnarray}
\label{delu}
\delta u(x) = 0.02\ (1-x)^{0.1},
\end{eqnarray}
is rather hard, as illustrated in Fig.1.
Kuhlmann {\em et al.} \cite{KLT} observed that through the QCD evolution
feed-down effect, evolving this component from $Q^2 \sim 25$~GeV$^2$ to
$Q^2 \sim 40000$~GeV$^2$ increases the standard model cross section 
significantly beyond $x=0.75$, without sacrificing the quality of the
global fits to existing data at lower energies.

While a distribution of type (\ref{delu}) can lead to better
agreement with the HERA data, the physical origin of such a hard
component of the $u$ quark distribution is difficult to motivate,
given the small mass of the $u$ quark.
In addition, it was argued in a recent reanalysis \cite{RB} of SLAC
data for $x > 0.7$ that the proposed modification \cite{KLT} would
overestimate the large-$x$ data by some two orders of magnitude.

Another possibility raised in Ref.\cite{KLT} is that the charm quark
distribution might be enhanced at large $x$.
Because the couplings of $u$ and $c$ quarks to electroweak bosons
are identical, the effect of a modified $c$ distribution would be
the same as that of a modified $u$.
In addition, one could more easily imagine that the charm quark distribution 
could be rather hard, owing to the large $c$ quark mass.
Such a component would of course have to be generated non-perturbatively,
since the charm quark distribution arising from gluon bremsstrahlung alone
looks like a typical soft sea distribution, namely peaks at $x \rightarrow 0$
and is negligible beyond $x \sim 0.4$, Fig.1.

The effect of a non-perturbative, or intrinsic, charm component on the
DIS cross sections at HERA kinematics was recently investigated by Gunion
and Vogt \cite{GV} within a model of the 5-quark component of the nucleon
wave function on the light-cone \cite{BROD}.
Following Brodsky {\em et al.} \cite{BROD}, the wave function was assumed
to be inversely proportional to the light-cone energy difference between
the nucleon ground state and the 5-quark excited state.
The resulting $x$-dependence of the inclusive $c$ quark distribution in the
minimal model of \cite{GV} was given by \cite{BROD}:
\begin{eqnarray}
\label{delic}
\delta^{(IC)} c(x)
&=& 6 x^2 \left( (1-x) (1 + 10x + x^2) - 6x (1+x) \log 1/x \right),
\end{eqnarray}
with normalization fixed to 1\% \cite{GV}.
As seen in Fig.1, such a distribution peaks at $x \sim 0.2$, and is
negligible beyond $x \sim 0.7$.
The anti-charm distribution is assumed to be equivalent to the charm
distribution in this model,
$\delta^{(IC)} \overline c(x) = \delta^{(IC)} c(x)$.
After evolving to the HERA kinematics, this intrinsic charm distribution,
while considerably harder than that generated through pQCD alone, is still
too soft to account for the excess HERA events \cite{GV}.

What one needs, therefore, is a substantially harder distribution which
has significantly more strength above $x \sim 0.6$ than in (\ref{delic}).
In this paper we investigate whether one can indeed obtain a sizable
enhancement of the number of events at large $x$ and $Q^2$ from a hard
charm distribution, and whether such a distribution can be motivated
from any physical model.

As an alternative to the intrinsic charm picture of Refs.\cite{GV,BROD},
the charmed sea was taken in Refs.\cite{NNNT,PNNDB} to arise from the
quantum fluctuation of the nucleon to a virtual $D^- + \Lambda_c$
configuration.
The nucleon charm radius \cite{NNNT} and the charm quark distribution 
\cite{PNNDB} were both estimated in this framework.
The meson cloud model for the long-range structure of the nucleon has been
developed in Refs.\cite{MCM,HSS,ST} to describe various flavor symmetry
breaking phenomena observed in DIS and related experiments.
It offers a natural explanation of the $\overline d$ excess in the proton
over $\overline u$ \cite{EXCESS} in terms of a pion cloud, which itself 
is a necessary ingredient of the nucleon by chiral symmetry.
It also provides an intuitive framework to study the strangeness content
of the nucleon, through the presence of the kaon cloud of the nucleon
\cite{HSS,STRANGE}.
Whether the same philosophy can be justified for a cloud of heavy charmed
mesons and baryons around the nucleon is rather more questionable given
the large mass of the fluctuation.
Nevertheless, to a crude approximation, we may take the meson cloud
framework as an indicator of the possible order of magnitude and shape
of the non-perturbative charm distribution.
Furthermore, a natural prediction of this model is non-symmetric $c$
and $\overline c$ distributions.

In the meson cloud model, the distribution of charm quarks in the
nucleon on the light cone at some low hadronic scale is written
in convolution form:
\begin{eqnarray}
\delta \overline c(x)
&=& \int_x^1 {dz \over z} f_{D/N}(z)\ 
	\overline c^{D^-}\left({x \over z}\right),\ \ \ \ 
\delta c(x)
\ =\ \int_x^1 {dz \over z} f_{\Lambda_c/N}(z)\ 
	c^{\Lambda_c}\left({x \over z}\right),
\end{eqnarray}
where $z$ is the fraction of the nucleon's light-cone momentum 
carried by the $D^-$ meson or $\Lambda_c$.
The light cone (or infinite momentum frame) distribution of $D^-$
mesons in the nucleon is given by:
\begin{eqnarray}
\label{fz}
f_{D/N}(z)
&=& { 1 \over 16 \pi^2 }
\int_0^\infty dk^2_\perp
{ g^2(k_\perp^2,z) \over z (1-z) (s_{D \Lambda_c} - M_N^2)^2 }
\left( { k_\perp^2 + [M_{\Lambda_c} - (1-z) M_N]^2 \over 1-z } \right),
\end{eqnarray}
and can be shown to be related to the light-cone distribution of 
$\Lambda_c$ baryons, $f_{\Lambda_c/N}(z)$,
by $f_{\Lambda_c/N}(z) = f_{D/N}(1-z)$.

In Eq.(\ref{fz}) the function $g$ describes the extended nature of the
$D\Lambda_c N$ vertex, with the momentum dependence parameterized by\
$g^2(k_\perp^2,z)\
=\ g_0^2\ (\Lambda^2 + M_N^2)/(\Lambda^2 + s_{D\Lambda_c})$,
where the $D\Lambda_c$ center of mass energy squared
is given by\ 
$s_{D \Lambda_c} = (k_\perp^2 + M_D^2)/z
		 + (k_\perp^2 + M_{\Lambda_c}^2)/(1-z)$,
and $g_0$ is the $D \Lambda_c N$ coupling constant at the
pole, $s_{D \Lambda_c} = M_N^2$.
We expect $g_0$ to be similar to the $\pi NN$ coupling constant.

Because of the large mass of the $c$ quark, one can approximate the
$\overline c$ distribution in the $D^-$ meson \cite{PNNDB} and the
$c$ distribution in the $\Lambda_c^+$ by:
\begin{eqnarray}
\overline c^{D^-}(x) &\approx& \delta(x-1),\ \ \ \ \
c^{\Lambda_c^+}(x)\ \approx\ \delta(x-2/3),
\end{eqnarray}
which then gives:
\begin{eqnarray}
\label{mcm_final}
\delta \overline c(x) &\approx& f_{D/N}(x),\ \ \ \ \ \ 
\delta c(x)\ \approx\ {3 \over 2} f_{\Lambda_c/N}(3x/2).
\end{eqnarray}

The resulting $\delta c$ and $\delta \overline c$ distributions are shown
in Fig.1, calculated for an ultraviolet cutoff of $\Lambda \approx 2.2$~GeV,
which gives
$\int_0^1 dx \delta c(x) = \int_0^1 dx \delta \overline c(x) \approx 1\%$.
For a probability of 0.5\% one would need a smaller cutoff,
$\Lambda \approx 1.7$~GeV.
Quite interestingly, the shape of the $c$ quark distributions is quite
similar to that in the intrinsic charm model of Refs.\cite{GV,BROD}.
However, as mentioned above, the model of \cite{GV,BROD} assumes identical
shapes for the non-perturbative $c$ and $\overline c$ distributions, while
the meson cloud gives a significantly harder $\overline c$ distribution.

Calculation of the NC and CC cross sections requires parton distributions
for all flavors.
For this we use a recent parameterization of global data from the CTEQ
Collaboration \cite{CTEQ}.
Expressions for the differential NC and CC cross sections $d^2\sigma/dxdQ^2$
in the standard model can be found in Refs.\cite{ZEUS} and \cite{H196}.
In Fig.2 we show the ratios of the modified to standard DIS model NC
cross sections, with $\sigma \equiv d^2\sigma/dxdQ^2$, and 
$\sigma + \delta\sigma$
represents the cross section calculated with the modified distributions.
The result with the modified $u$ distribution \cite{KLT} rises sharply
above $M \sim 200$ GeV.
The effect is rather similar if one includes the non-perturbative $\delta c$
and $\delta \overline c$ distributions from the meson cloud model.
On the other hand, with the somewhat softer, intrinsic charm distribution
of \cite{GV,BROD}, the enhancement is rather modest, and $\alt 10\%$ over
the whole range of $M$.

Also shown in Fig.2 is the effect of additional contributions to the light
flavor distributions ($u, \overline u, d$ and $\overline d$) which one would
obtain from a pion cloud of the nucleon \cite{MCM}.
Although these are considerably larger in magnitude than the $\delta c$
or $\delta \overline c$ distributions, because they appear at small $x$
($\sim 0.1$) their effect is to yield only a very small enhancement of 
the cross section ratio.
{}From this figure one can conclude that the only realistic candidates
for a significant enhancement of the cross section at large $M$ are the
modified valence $u$ distribution from Ref.\cite{KLT}, and the hard charm
distributions in Eq.(\ref{mcm_final}).

To compare with the H1 and ZEUS data one needs to evolve the cross sections
to the typical scales relevant for HERA.
Since the excess of events is seen at quite large $Q^2$, 
$10000 \alt Q^2 \alt 40000$~GeV$^2$, we evolve the $\delta u$ and 
$\delta c$ \& $\delta \overline c$ distributions to an average value
of $Q^2 = 20000$~GeV$^2$.
The result is shown in Fig.3, where for $M \agt 250$~GeV especially one
has an enormous enhancement of the NC cross section.
The enhancement is qualitatively similar for both the 0.5\% and 1\%
$\delta \overline c$ components.

For scattering via the CC, the effect of the additional contributions to
the parton distributions is shown in Fig.4 for $Q^2 = 20000$~GeV$^2$.
Since the $W^+$ boson is not sensitive to the $u$ quark in the proton,
the $\delta u$ modification has no effect on the $e^+ p$ cross section.
In contrast, as noted by Babu {\it et al.} \cite{BABU}, the effect of
an additional non-perturbative $\delta \overline c$ contribution is an
even larger enhancement of the CC cross section than the NC cross section.
With a 0.5\% (1\%) intrinsic charm component the CC cross section
increases by a factor $\sim$ 2 (3) for $200 \alt M \alt 250$~GeV,
which is similar to the excess observed by H1 \cite{H1} in this region.

Having seen that one can indeed modify the existing parton distributions
at large $x$ to produce large enhancements of the cross sections at large
$M$ and $Q^2$, the key question to address next is whether such modified
distributions are compatible with existing data at lower $Q^2$.
A recent analysis \cite{RB} of SLAC data at $0.7 \leq x \leq 0.97$ in
and near the resonance region suggests that the proposed modification
of the $u$ quark distribution in Ref.\cite{KLT} would overestimate
the large-$x$ data by two orders of magnitude.
(Though one should keep in mind that the authors of Ref.\cite{RB} 
assume the validity
of Bloom-Gilman duality even at very large $x$ values close to 1.)
On the other hand, the values of $W^2$ corresponding to the $x$ and 
$Q^2$ bins are too low for the SLAC data to be sensitive to the 
charm component of the nucleon wave function.
Data are, however, available from the BCDMS Collaboration in the region
$x > 0.6$ which are above charm threshold \cite{BCDMS}.
The effect of adding the intrinsic $\delta \overline c$ distribution
calculated from Eq.(\ref{mcm_final}) to the data on the deuteron structure
function (as parameterized in Ref.\cite{NMC}) is illustrated in Fig.5.
With the addition of the 0.5\% contribution there does not seem to be any
inconsistency with the data \cite{BCDMS} and, indeed, the agreement is
slightly improved at $x=0.75$.
On the other hand, the 1\% case may be a little too high for comfort.
Earlier data from the EMC \cite{EMC} on the charm structure function
extracted from charmed particle production in $\mu$--nucleus scattering
also exist, but cover only the region below $x \sim 0.3$.

One can conclude, therefore, that a non-perturbatively generated hard
charm distribution can significantly enhance the DIS cross sections
at the large $x$ and $Q^2$ values attainable at HERA, and still be
compatible with data at lower $Q^2$.
We should make clear, however, that the $\delta c$ and $\delta \overline c$
quark distributions calculated in the meson cloud model should not be viewed
as quantitative predictions, but rather as a physical motivation for a hard
non-perturbative charm component of the nucleon.
In particular, one can vary model parameters to modify the detailed shape
of the distributions.
For example, using a smaller power of $1/(\Lambda^2+s_{D\Lambda_c})$
in the hadronic vertex function $g(k_\perp^2,z)$ would result in a
$\delta \overline c$ distribution with even more strength at $x > 0.75$
than that in Fig.1, which would further enhance the cross sections for
$M \sim 200 - 250$ GeV.
Nevertheless, the fact that the model predicts a large asymmetry between
$c$ and $\overline c$ distributions is a robust, and to our knowledge
unique, feature of the meson cloud model.

Another possibility which could lead to additional enhancement of the
cross sections at large $x$ would be a non-perturbative $\overline b$
quark component \cite{KLT}, although the magnitude of this would
presumably be rather small compared with the $\overline c$.
In addition, the recent reanalysis of the SLAC deuteron data in
Ref.\cite{MT} suggested that the valence $d/u$ ratio does not tend
to zero as $x \rightarrow 1$, as assumed in global fits to data
\cite{CTEQ}, but rather is consistent with the expectation of
perturbative QCD, namely $d/u \rightarrow 1/5$ \cite{FJ}.
Figure 6 shows the ratio of the modified $d$ quark distribution,
$d + \delta d$, extracted from the deuteron and proton data in
Ref.\cite{MT} at $Q^2 \approx 10$ GeV$^2$ and evolved to
$Q^2 = 10000$ GeV$^2$, to that used in the standard fits.
The correction $\delta d$ is taken from the parameterization in
Ref.\cite{MP}.
The effect of the modified $d$ quark is not large in NC events,
primarily due to the factor 4 charge suppression compared with $u$,
but can be sizable for CC events, comparable to that of the 0.5\%
$\delta \overline c$ scenario --- as shown by the dashed curve in Fig.4.
The fact that all standard sets of parton distributions assume that
$d/u \rightarrow 0$ as $x \rightarrow 1$ means that there is a possible
source of systematic error in the modeling of ``background'' events
which should be accounted for.

The task of disentangling the origin of the HERA anomaly in future
can be approached from several directions.
Tagging charm final states in $J/\psi$ production could provide information
on whether the large $x$ enhancement is due to charm or other flavors.
A thorough comparison of NC and CC cross sections, for both $e^+ p$
and $e^- p$ collisions, would also enable one to determine the flavor
combinations responsible for the cross section enhancement \cite{BABU}.
One can also construct new distributions, such as those suggested in
Ref.\cite{MCK}, which are insensitive to parton distributions, but sensitive
to modifications of parton level cross sections, and which could therefore
distinguish between explanations of the anomaly in terms of modifications
of parton distributions and those invoking new physics.

Needless to say, we eagerly await further results from the H1 and ZEUS
Collaborations.
Quite apart from the excitement over the possible discovery of physics
beyond the standard model, the issue of whether or not the nucleon
contains a significant component of intrinsic charm is extremely important.
In particular, any experimental evidence supporting the suggestion that
the intrinsic charm could have a strong asymmetry would mean a revision
of current wisdom, and would undoubtedly lead us to a deeper understanding 
of the structure of hadronic systems.
Whether H1 and ZEUS provide evidence for new physics beyond the standard
model or map out hitherto unknown features of the non-perturbative structure
of the nucleon, the results will be of great interest.

%%%%%%%%%%%%%%%%%%%%%%%%%%%%%%%%%%%%%%%%%%%%%%%%%%%%%%%%%%%%%%%%%%%%%%%%%%%%%%%%
\acknowledgements

W.M. would like to thank the Special Research Centre for the Subatomic
Structure of Matter at the University of Adelaide for support during a
recent visit, where this work was initiated.
This work was supported by the DOE grant DE-FG02-93ER-40762 and by the 
Australian Research Council.

%%%%%%%%%%%%%%%%%%%%%%%%%%%%%%%%%%%%%%%%%%%%%%%%%%%%%%%%%%%%%%%%%%%%%%%%%%%%%%%%

%%%%%%%%%%%%%%%%%%%%%%%%%%%%%%%%%%%%%%%%%%%%%%%%%%%%%%%%%%%%%%%%%%%%%%%%%%
\begin{figure}
\label{fig1}
\epsfig{figure=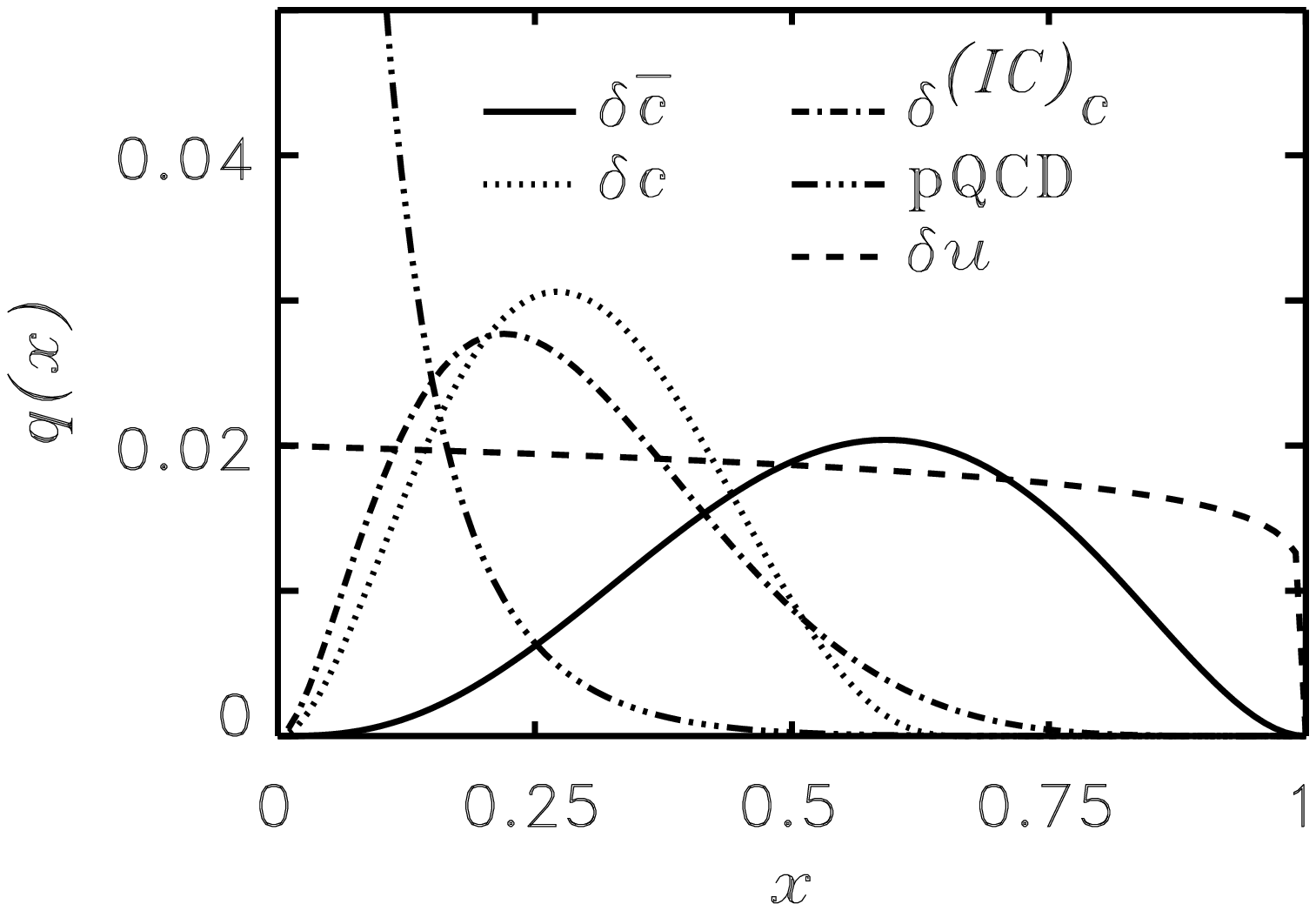,height=8.5cm}
\caption{Modification $\delta u$ of the valence $u$ distribution from
	Ref.\protect\cite{KLT}, compared with the light-cone intrinsic
	charm distribution, $\delta^{(IC)} c$, of Refs.\protect\cite{GV,BROD},
	and the non-perturbative $\delta c$ and $\delta \overline c$
	distributions in the meson cloud model (the latter normalized to 1\%).
	Also shown is the purely perturbative contribution at $Q^2=$4~GeV$^2$.}
\end{figure}

\begin{figure}
\label{fig2}
\epsfig{figure=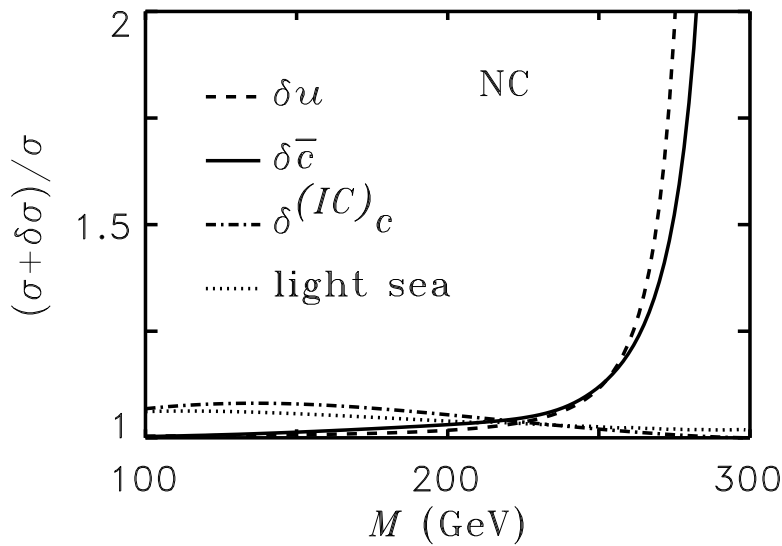,height=8.5cm}
\caption{Ratio of modified to standard DIS model NC cross sections
	at the charm threshold, with the modifications arising from
	the additional $u$ quark component \protect\cite{KLT} (dashed),
	1\% non-perturbative $\delta \overline c$ and $\delta c$
	distributions from the meson cloud model (solid),
	and the intrinsic charm model of Refs.\protect\cite{GV,BROD}
	(dot-dashed).
	Also shown is the effect of the meson cloud contributions to
	the light sea quarks \protect\cite{MCM} (dotted).}
\end{figure}

\begin{figure}
\label{fig3}
\epsfig{figure=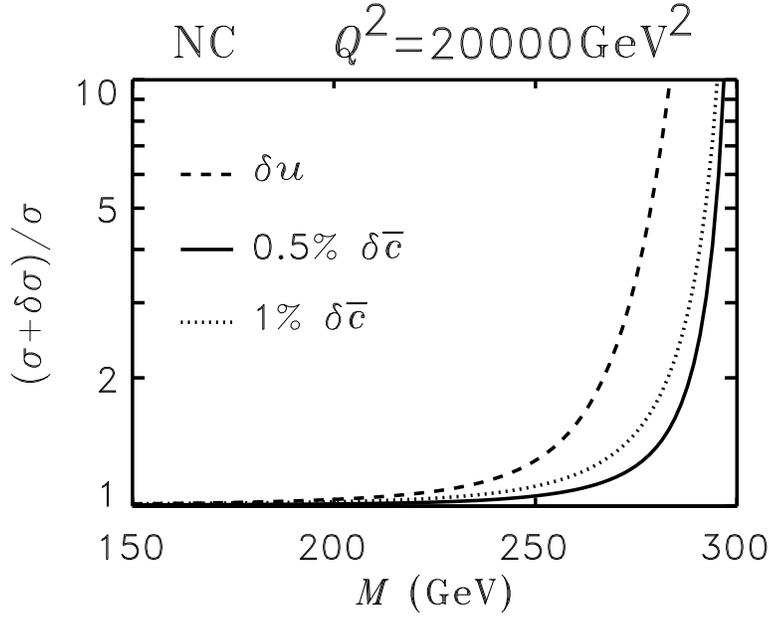,height=8.5cm}
\caption{Ratio of modified to standard DIS model NC cross sections at
	$Q^2=20000$~GeV$^2$.
	The modifications are from the extra $\delta u$ component
	\protect\cite{KLT}, and from additional hard non-perturbative
	$\delta c$ and $\delta \overline c$ components, with 0.5\% and
	1\% normalization.}
\end{figure}

\begin{figure}
\label{fig4}
\epsfig{figure=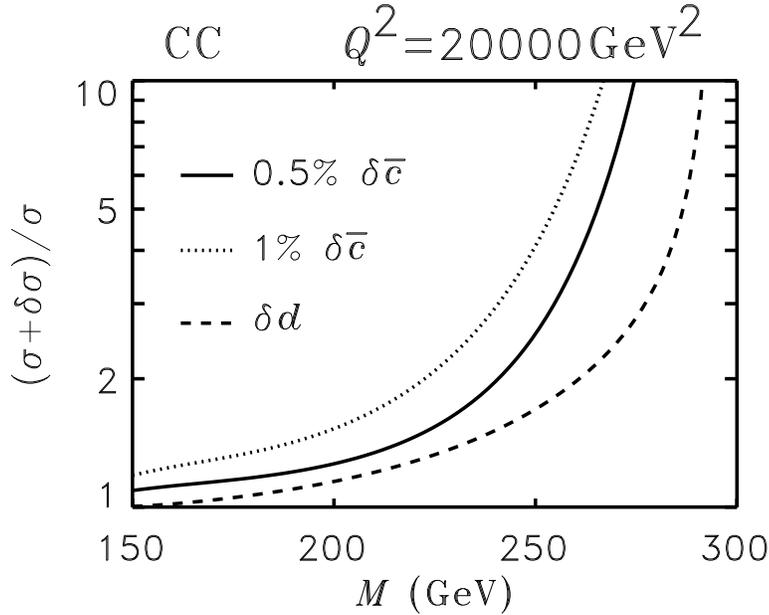,height=8.5cm}
\caption{Ratio of modified to standard DIS model CC cross sections at
	$Q^2=20000$~GeV$^2$, with the modifications arising from 
	0.5\% and 1\% additional $\delta \overline c$ distributions,
	as well as a modified $d$ quark distribution at large $x$
	(see Fig.6).}
\end{figure}

\begin{figure}
\label{fig5}
\epsfig{figure=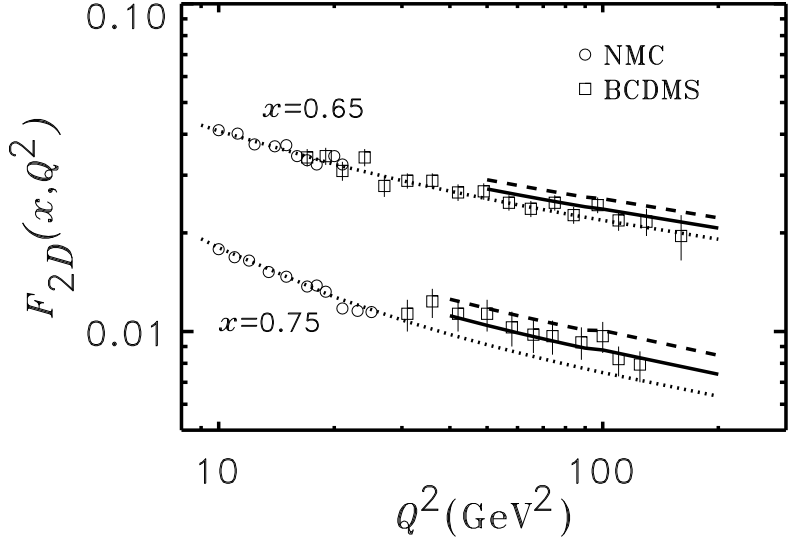,height=8.5cm}
\caption{Structure function of the deuteron, $F_{2D}$, with 0.5\% (solid)
	and 1\% (dashed) additions of non-perturbative charm to the global
	fit (dotted) from Ref.\protect\cite{NMC}.}
\end{figure}

\begin{figure}
\label{fig6}
\epsfig{figure=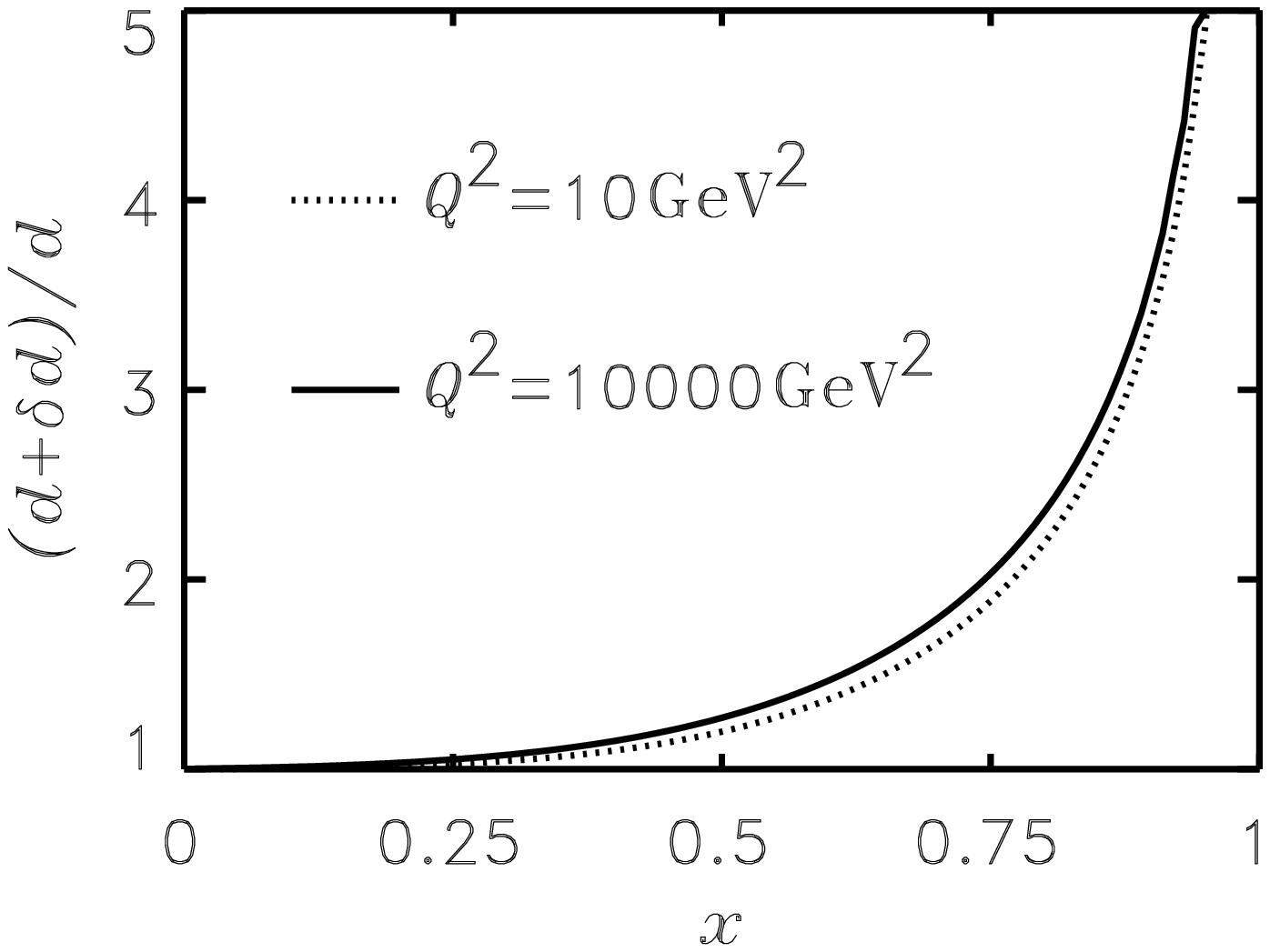,height=8.5cm}
\caption{Ratio of the modified $d$ quark distribution, $d+\delta d$,
	extracted from deuteron data in Ref.\protect\cite{MT} at
	$Q^2 \approx 10$ GeV$^2$ and evolved to $Q^2 = 10000$ GeV$^2$,
	to that used in standard, global fits \protect\cite{CTEQ},
	which assume $d/u \rightarrow 0$ as $x \rightarrow 1$.}
\end{figure}


\references

\bibitem{H1}
H1 Collaboration, C. Adloff {\em et al.},
DESY 97-024 [hep-ex/9702012].

\bibitem{ZEUS}
ZEUS Collaboration, J. Breitweg {\em et al.},
DESY 97-025 [hep-ex/9702015].

\bibitem{REV}
Yu.L. Dokshitzer,
to appear in the Proceedings of the 5th International Workshop
on DIS and QCD, Chicago, April 1997 [hep-ph/9706375];
%
G. Altarelli,
CERN-TH-97-195 [hep-ph/9708437].

\bibitem{COMP}
G. Altarelli, J. Ellis, G.F. Giudice, S. Lola and M. Mangano,
CERN-TH-97-40 [hep-ph/9703276];
%
M. Drees,
Phys. Lett. B {\bf 403}, 353 (1997);
%
U. Bassler and G. Bernardi,
DESY 97-136 [hep-ex/9707024].

\bibitem{KLT}
S. Kuhlmann, H.L. Lai and W.K. Tung,
MSU-HEP-70316, CTEQ-705 [hep-ph/9704338].

\bibitem{RB}
S. Rock and P. Bosted,
American University report [hep-ph/9706436].

\bibitem{GV}
J.F. Gunion and R. Vogt,
UCD-97-14, LBNL-40399 [hep-ph/9706252].

\bibitem{BROD}
S.J. Brodsky, P. Hoyer, C. Peterson and N. Sakai,
Phys. Lett. {\bf 93} B, 451 (1980);
%
S.J. Brodsky, C. Peterson and N. Sakai,
Phys. Rev. D {\bf 23}, 2745 (1981).

\bibitem{NNNT}
F.S. Navarra, M. Nielsen, C.A.A. Nunes and M. Teixeira,
Phys. Rev. D {\bf 54}, 842 (1996).

\bibitem{PNNDB}
S. Paiva, M. Nielsen, F.S. Navarra, F.O. Duraes and L.L. Barz,
IFUSP-P-1240 [hep-ph/9610310].

\bibitem{MCM}
A.W. Thomas,
Phys. Lett. {\bf 126} B, 97 (1983);
%
W. Melnitchouk, A.W. Thomas and A.I. Signal,
Z. Phys. A {\bf 340}, 85 (1991);
%
V.R. Zoller, Z. Phys. C {\bf 60}, 141 (1993);
%
W. Melnitchouk and A.W. Thomas,
Phys. Rev. D {\bf 47}, 3794 (1993).

\bibitem{HSS}
H. Holtmann, A. Szczurek and J. Speth,
Nucl. Phys. {\bf A569}, 631 (1996).

\bibitem{ST}
J. Speth and A.W. Thomas,
TJNAF report PRINT-96-213 (1996),
to appear in Adv. Nucl. Phys.

\bibitem{EXCESS}
New Muon Collaboration, P. Amandruz {\em et al.},
Phys. Rev. Lett. {\bf 66}, 2712 (1993);
%
NA51 Collaboration, A. Baldit {\em et al.},
Phys. Lett. B {\bf 332}, 244 (1994);
%
Fermilab E866 Collaboration, P.L. McGaughey {\em et al.},
to be published.

\bibitem{STRANGE}
A.I. Signal and A.W. Thomas,
Phys. Lett. {\bf 191} B, 205 (1987);
%
M.J. Musolf {\em et al.},
Phys. Rep. {\bf 239}, 1 (1994);
%
W. Melnitchouk and M. Malheiro,
Phys. Rev. C {\bf 55}, 431 (1997).

\bibitem{CTEQ}
H.L. Lai, J. Huston, S. Kuhlmann, F. Olness, J.F. Owens,
D. Soper, W.K. Tung and H. Weerts,
Phys. Rev. D {\bf 55}, 1280 (1997).

\bibitem{H196}
H1 Collaboration, S. Aid {\em et al.},
Phys. Lett. B {\bf 379}, 319 (1996).

\bibitem{BABU}
K.S. Babu, C. Kolda and J. March-Russell,
IASSNS-HEP-97-55 [hep-ph/9705399].

\bibitem{BCDMS}
BCDMS Collaboration, A.C. Benvenuti {\em et al.},
Phys. Lett. B {\bf 223} 485, (1989);
%
BCDMS Collaboration, A.C. Benvenuti {\em et al.},
Phys. Lett. B {\bf 237} 592, (1990).

\bibitem{NMC}
New Muon Collaboration, M. Arneodo {\em et al.},
Phys. Lett. B {\bf 364} 107, (1995).

\bibitem{EMC}
European Muon Collaboration, J.J. Aubert {\em et al.},
Phys. Lett. {\bf 110B}, 73 (1982).
%
Nucl. Phys. {\bf B213} 213, (1983);

\bibitem{MT}
W. Melnitchouk and A.W. Thomas,
Phys. Lett. B {\bf 377}, 11 (1996).

\bibitem{FJ}
G.R. Farrar and D.R. Jackson,
Phys. Rev. Lett. {\bf 35}, 1416 (1975).

\bibitem{MP}
W. Melnitchouk and J.-C. Peng,
Phys. Lett. B {\bf 400}, 220 (1997).

\bibitem{MCK}
Z. Cao, X.-G. He and B. McKellar,
OITS-633 [hep-ph/9707227].
\end{document}